\documentclass{article} %
\usepackage[final,main]{format/neurips_2026}

\usepackage{microtype}
\usepackage{hyperref}
\usepackage{url}
\usepackage{booktabs}
\usepackage{multirow}
\usepackage{colortbl}
\usepackage{xspace}
\usepackage{enumitem} 
\usepackage{xcolor}   
\usepackage{url}     
\usepackage{amsmath} 
\usepackage{amssymb}
\usepackage{amsthm}
\usepackage{wrapfig}
\usepackage{graphicx}
\usepackage{pifont}
\usepackage{tcolorbox}
\tcbuselibrary{skins,breakable}
\usepackage{listings}
\usepackage{algorithm}
\usepackage{algpseudocode}
\usepackage{nicefrac}
\usepackage{tabularx}
\usepackage{array}

\definecolor{promptbg}{RGB}{245,245,245}
\definecolor{promptframe}{RGB}{180,180,180}
\definecolor{outputbg}{RGB}{240,247,255}
\definecolor{outputframe}{RGB}{140,170,210}
\definecolor{codekeyword}{RGB}{0,92,175}
\definecolor{codeverus}{RGB}{125,70,155}
\definecolor{codecomment}{RGB}{95,110,120}
\definecolor{codetype}{RGB}{0,120,120}
\definecolor{codenumber}{RGB}{130,130,130}

\newtcolorbox{promptbox}[1][]{
  colback=promptbg, colframe=promptframe,
  fonttitle=\bfseries\small, title={#1},
  boxrule=0.4pt, arc=1.5pt, left=5pt, right=5pt, top=3pt, bottom=3pt,
  breakable
}

\newtcolorbox{outputbox}[1][]{
  colback=outputbg, colframe=outputframe,
  fonttitle=\bfseries\small, title={#1},
  boxrule=0.4pt, arc=1.5pt, left=5pt, right=5pt, top=3pt, bottom=3pt,
  breakable
}

\lstdefinestyle{verusstyle}{
  basicstyle=\ttfamily\scriptsize,
  keywordstyle=\color{codekeyword}\bfseries,
  keywordstyle=[2]\color{codeverus}\bfseries,
  commentstyle=\color{codecomment}\itshape,
  emphstyle=\color{codetype},
  emph={Vec,i32,int,bool,usize},
  numbers=left,
  numberstyle=\tiny\color{codenumber},
  numbersep=5pt,
  xleftmargin=1.6em,
  breaklines=true,
  frame=none,
  columns=fullflexible,
  keepspaces=true,
  aboveskip=2pt, belowskip=2pt
}

\lstdefinelanguage{Rust}{
  morekeywords={fn,let,mut,while,if,else,return,pub,use,as},
  morekeywords=[2]{forall,invariant,requires,ensures,decreases,assert,proof,by,open,spec},
  sensitive=true,
  morecomment=[l]{//},
}

\usepackage{lineno}

\definecolor{darkblue}{rgb}{0, 0, 0.5}
\hypersetup{colorlinks=true, citecolor=darkblue, linkcolor=darkblue, urlcolor=darkblue}

\usepackage{comment}

\newcommand{\tool}{\textsc{TopoVerus}\xspace}

\newcommand{\ie}{i.e., }
\newcommand{\eg}{e.g., }
\newcommand{\etc}{etc.\xspace}

\newcommand{\para}[1]{\smallskip\noindent{\textbf{#1}}}

\definecolor{blue-green}{rgb}{0.0, 0.87, 0.87}

\newtheorem{definition}{Definition}
\theoremstyle{remark} 
\newtheorem*{remark}{Remark}

\lstdefinestyle{casealign}{xleftmargin=3.4pt,xrightmargin=3.4pt,aboveskip=0pt,belowskip=0pt}
\newsavebox{\caseleft}\newsavebox{\caserighttop}\newsavebox{\caserightbot}
\newlength{\caseleftw}\newlength{\caserightw}
\newcommand{\casehead}[1]{\textbf{\small #1}\par\vspace{2pt}}
\newcommand{\casetotal}[1]{\dimexpr\ht#1+\dp#1\relax}
\newsavebox{\casectx}\newlength{\casegap}\newlength{\casectxht}
\newcommand{\casesep}{\par\vspace{\casegap}}

\definecolor{caseblue}{RGB}{34,74,126}
\definecolor{casegreen}{RGB}{32,104,92}
\definecolor{casered}{RGB}{164,55,55}
\definecolor{casegray}{RGB}{90,90,90}
\definecolor{casecodebg}{RGB}{248,248,246}
\definecolor{casectxbg}{RGB}{246,250,252}
\definecolor{caseerrorbg}{RGB}{255,248,246}
\definecolor{caseframe}{RGB}{205,211,218}

\lstdefinestyle{caseverus}{
  style=verusstyle,
  basicstyle=\ttfamily\tiny,
  keywordstyle=\color{caseblue}\bfseries,
  commentstyle=\color{casegray},
  backgroundcolor=\color{casecodebg},
  frame=single,
  rulecolor=\color{caseframe},
  xleftmargin=2pt,
  xrightmargin=2pt,
  aboveskip=2pt,
  belowskip=2pt,
  numbers=none
}

\lstdefinestyle{caseerror}{
  basicstyle=\ttfamily\tiny,
  breaklines=true,
  frame=single,
  columns=fullflexible,
  keepspaces=true,
  backgroundcolor=\color{caseerrorbg},
  rulecolor=\color{caseframe},
  xleftmargin=2pt,
  xrightmargin=2pt,
  aboveskip=2pt,
  belowskip=2pt,
  numbers=none
}

\newtcolorbox{casebox}[2][]{
  enhanced,
  colback=white,
  colframe=caseframe,
  coltitle=black,
  fonttitle=\bfseries\small,
  title={#2},
  boxrule=0.45pt,
  arc=1.2pt,
  left=4pt,
  right=4pt,
  top=4pt,
  bottom=4pt,
  #1
}

\newtcolorbox{ctxbox}[1][]{
  enhanced,
  colback=casectxbg,
  colframe=caseframe,
  boxrule=0.35pt,
  arc=1.2pt,
  left=4pt,
  right=4pt,
  top=3pt,
  bottom=3pt,
  fontupper=\scriptsize,
  #1
}

\newcommand{\caseheader}[2]{
  \noindent\textbf{Task.} \texttt{#1}\\[-1pt]
  \textit{#2}
  \vspace{3pt}
}

\title{Curriculum Proof Repair with\\ Learned Counterexamples}

\author{
Yuechun Sun \quad Jun Yang \quad Kexin Pei \\
Department of Computer Science \\
University of Chicago \\
Chicago, IL 60637, USA \\
\texttt{\{ycsun,juny,kpei\}@uchicago.edu}
}

\begin{document}
\maketitle

\begin{abstract}
Language models (LMs) have substantially lowered the barrier to applying formal verification for building reliable and secure software.
Existing approaches commonly follow the verifier-in-the-loop paradigm, but the verifier often provides only coarse, binary pass/fail feedback, which offers little actionable signal on how the proof should be repaired.
Recent work shows that counterexamples provide more actionable feedback by grounding proof failures in concrete program states, but generating useful ones at inference time requires costly iterative refinement.
We present \tool, an LM-based framework for Verus proof repair that internalizes counterexample-guided diagnosis as a model capability, producing useful counterexamples in a single pass.
Our key insight is to train counterexample generation not as a standalone validation task, but as a data-driven diagnostic objective: a counterexample is useful when conditioning on it helps the model produce a safe, verifier-passing proof. 
Given a failed Verus proof and verifier error, \tool generates a structured source-level counterexample, repair rationale, and repaired proof in one trajectory, optimized with reinforcement learning from Verus feedback and a proof-code safety gate. 
To guide the order of invariant repairs and target counterexamples to specific invariants, \tool introduces a curriculum with a topological graph reward based on invariant dependencies derived from inductiveness tests and encourages counterexample generation following the same topological order as these dependencies. 
Across seven Verus benchmarks with 471 problems, \tool achieves 61.6\%/72.2\% weighted-average Safe-Pass@1/Safe-Pass@3, outperforming the strongest baseline, Claude Sonnet 4.5, by 5.7/6.3 points.
We also show that \tool's learned counterexamples can directly augment the reasoning of existing LM-based proof repair approaches, improving their success rate by 64.3\% compared to chain-of-thought diagnostics. 
We release the training and evaluation code at \url{https://github.com/Etha-Sun/Topoverus} and the \tool-32B weights at \url{https://huggingface.co/YuechunSun/TopoVerus-32B}.
\end{abstract}

\section{Introduction}

Formal verification provides machine-checked guarantees that software satisfies rich correctness specifications, enabling the construction of safety-critical, large-scale software systems~\citep{lattuada2023verus,leroy2016compcert, klein2009sel4, anvil, verismo, microkernel}. 
However, these guarantees remain costly to obtain.
Developers must write specifications, discover inductive invariants, and repair failed proof obligations. 
Recent work using language models (LMs) for verification automates parts of this proof-construction loop~\citep{sun2026veristruct, zhong2025rag, Chen2025Automated, yang2024autoverus, yang2026verusage, zhong2025towards, Aggarwal2025AlphaVerus, mugnier2025laurel, first2023baldur, yang2026exverus, li2026p, ye2026vero}, lowering the expertise barrier for applying formal verification to real-world software.

As LM-generated proof candidates are not sound by construction, existing LM-based approaches typically keep a verifier in the loop: the model proposes proof annotations, the verifier reports failures, and the model revises the proof based on the error messages~\citep{yang2024autoverus, sun2026veristruct, zhong2025rag, Aggarwal2025AlphaVerus}. 
This loop resembles interactive verification, where developers debug failures by constructing concrete failing states and tracing where an invariant or postcondition breaks~\citep{clarke2000counterexample, clarke2003counterexample}. 
However, verifier messages often expose only a failed proof obligation with limited diagnostic context, e.g., \texttt{postcondition not satisfied}, rather than the source-level state, execution path, or missing facts responsible for the failure. 
The LM must thus infer the root cause from ambiguous feedback, making repair brittle and token-inefficient.

Existing LM-based proof repair approaches compensate for coarse verifier feedback in different ways.
Some encode verifier-specific repair rules or syntax guidance in the prompt~\citep{yang2024autoverus, sun2026veristruct}, supply in-context proof examples~\citep{yang2024autoverus, mugnier2025laurel}, or retrieve relevant proof fragments~\citep{zhong2025rag}. 
Others bypass explicit diagnosis and use verification outcomes, i.e., pass/fail signals, to guide search, data synthesis, self-improvement, or fine-tuning~\citep{Aggarwal2025AlphaVerus, Chen2025Automated}. 
While effective, these approaches leave a diagnostic gap: engineered prompt knowledge is costly to curate and does not generalize as verifier behavior and proof idioms evolve~\citep{yang2026verusage, yang2026exverus}, whereas verifier-outcome objectives provide only sparse, indirect signals for repair. 

A more actionable feedback signal is a source-level counterexample: a concrete program state, optionally with a short execution trace, that falsifies a proof obligation. 
Unlike verifier-specific repair rules, counterexamples describe the concrete failing program state, making them largely independent of diagnostic formats and directly useful for localizing repairs. 
They are central to classic counterexample-guided verification~\citep{clarke2000counterexample, clarke2003counterexample, bradley2011sat, bradley2012understanding}, where failed states drive refinement or blocking of bad regions. 
Recent LM-based works also show that generated counterexamples can improve proof repair~\citep{yang2026exverus, li2026learning, huang2026reinforcement}. 
Unfortunately, a useful counterexample requires nontrivial, execution-level code reasoning: it must be feasible, expose the relevant corner case, and identify the missing proof fact, rather than merely state a plausible failure~\citep{copet2025cwm}. 
This also explains why existing approaches often incur high inference costs, as they require iterative refinement to generate high-quality counterexamples~\citep{yang2026exverus}.
Moreover, proof failures often conflate multiple missing or incorrect invariants, so which counterexample is most useful to address first is not well defined.

\para{Our approach.} 
We present \tool, an LM-based verification framework for Verus proof repair that internalizes counterexample-guided diagnosis as a model capability. 
Our idea is to formulate counterexample generation as a diagnostic learning problem where the counterexample is useful not merely when it is plausible~\citep{yang2026exverus}, but when conditioning on it helps the model produce a safe, verifier-passing proof. 
Given a failed Verus proof attempt and the corresponding verifier error, \tool generates a structured source-level counterexample, a concise repair rationale, and a candidate repaired proof in this order within a single trajectory. 
We optimize the full trajectory using reinforcement learning with verifiable rewards (RLVR) from Verus, together with a proof-code safety gate that rejects repairs modifying the executable program or user-supplied specifications. 
As a result, counterexamples are not trained as standalone artifacts with hand-written validators~\citep{yang2026exverus} or predefined verifier-specific repair taxonomies~\citep{yang2024autoverus}. 
Instead, they receive indirect credit through downstream repair success, biasing the model toward diagnostics that are actionable for safe proof repair.

This formulation, however, creates a key technical challenge.
The training signal for counterexample utility is sparse, delayed, and order-sensitive. 
A failed proof often contains multiple missing or incorrect invariants, so final verifier success alone does not reveal whether a generated counterexample was wrong, merely unhelpful, or aimed at a downstream proof obligation whose preconditions were not yet established (Section~\ref{subsec:motivate}). 
To expose this latent structure during learning, \tool approximates dependencies among invariants with a graph derived from inductiveness tests, and rewards repaired proofs whose invariants respect a topological order over the resulting graph. 
The resulting topological graph reward turns sparse outcome feedback into a dependency-aware \emph{curriculum}, i.e., it establishes state-constraining facts first, then generates counterexamples and repairs for dependent obligations in the corresponding order, where the counterexample's diagnostics become informative.

\para{Results.} 
Our evaluation shows that \tool improves Verus proof repair and provides actionable diagnostics for downstream repair tools. 
Across seven Verus benchmarks comprising 471 problems, \tool achieves 61.6\% and 72.2\% in Safe-Pass@1 and Safe-Pass@3, outperforming the strongest baseline, i.e., Claude Sonnet 4.5, by \textbf{5.7} and \textbf{6.3} points, respectively. 
The gains are largest on structurally difficult settings such as VeriCogen and ObfsBench, where proof templates and syntactic hints are less reliable. 
Ablations show that counterexample-augmented reasoning in \tool improves weighted Safe-Pass@1 by 11.2 points over the repair with natural-language diagnostics, while the topological graph reward further improves over binary verifier-only RL by \textbf{7.9} points. 
Importantly, \tool's generated counterexamples also transfer to other LMs used as proof repairers, improving their repair success by up to 64.3\% (relative) over chain-of-thought diagnostics.
We release the training and evaluation code at \url{https://github.com/Etha-Sun/Topoverus} and the \tool-32B weights at \url{https://huggingface.co/YuechunSun/TopoVerus-32B}.

\section{Overview}
\label{sec:overview}

\begin{wrapfigure}[18]{r}{0.55\textwidth}
    \centering
    \begin{minipage}{0.55\textwidth}
    \vspace{-1.1cm}
    \begin{lstlisting}[
      style=verusstyle,
      language=Rust,
      basicstyle=\ttfamily\tiny,
      breaklines=true,
      breakatwhitespace=false
    ]
fn search_insert(nums: Vec<i32>, target: i32) -> (res: i32)
requires
  1 <= nums@.len() <= 10000,
  forall|i, j| 0 <= i < j < nums@.len() ==> nums@[i] < nums@[j],
ensures /* insertion-position postcondition */
  0 <= res <= nums@.len(),
  nums@[0] <= target <= nums@[nums@.len() - 1] ==>
    (forall|i| 0 <= i < res ==> nums@[i] < target) &&
    (forall|i| res <= i < nums@.len() ==> nums@[i] >= target),
{
  let mut low = 0i32;
  let mut high = nums.len() as i32 - 1;
  while low <= high
    invariant
      1 <= nums@.len() <= 10000,     //I1           
      0 <= low <= nums@.len(),       //I2  
      -1 <= high <= nums@.len() - 1, //I3        
      forall|i| 0 <= i < low ==> nums@[i] < target,          //I4
      forall|i| high < i < nums@.len() ==> nums@[i] > target //I5
  {
    let mid = ((low + high) as usize / 2) as i32;
    if nums[mid as usize] == target { return mid; }
    if target < nums[mid as usize] { high = mid - 1; }
    else { low = mid + 1; }
  }
  low
}
    \end{lstlisting}
    \end{minipage}
    \vspace{-6pt}
    \caption{Proof annotation for binary-search insertion.}
    \label{fig:motivating_code}
    \vspace{-1pt}
\end{wrapfigure}

We focus on Verus, a Rust-native verifier for verifying Rust programs.
A Verus proof consists of executable Rust code together with specifications (\eg preconditions and postconditions), and proof annotations (\eg invariants, assertions).
The verifier discharges the proof into Verification Conditions (VCs) and checks them with off-the-shelf SMT solvers~\citep{z3,cvc5}.
Among the proof annotations, inferring invariants has been a key challenge of automating formal verification~\citep{houdini,garg2014ice,kamath2023finding}.
Loop invariants are properties that hold at the beginning of every loop iteration.

\subsection{Motivating Example}
\label{subsec:motivate}

Figure~\ref{fig:motivating_code} shows the Verus code and proof that motivate our approach.
The function implements binary-search insertion: it returns the position of \texttt{target} in a sorted array, or the index where \texttt{target} should be inserted.
The \texttt{requires} (preconditions) assume the input to be a sorted array, and the \texttt{ensures} (postconditions) specify that \texttt{target} is inserted correctly.
The loop invariants are the properties that are derived from the preconditions and execution code.
They hold before entering the loop (base case) and remain true in each loop iteration (inductiveness).
Note that invariants often depend on other invariants
or are mutually dependent.
We use this example to illustrate such dependencies and how fixing invariants following such dependencies benefits counterexample reasoning, diagnosis and repair of buggy proofs.

\para{Semantic dependencies among invariants.}
The five invariant clauses in Figure~\ref{fig:motivating_code} play different semantic roles in verifying the program.
\textbf{I1} bounds the length range of \texttt{nums}.
\textbf{I2} and \textbf{I3} use that length fact to keep \texttt{low} and \texttt{high} within index ranges.
\textbf{I4} and \textbf{I5} are the semantic partition invariants: all indices strictly left of \texttt{low} contain values below \texttt{target}, while all indices strictly right of \texttt{high} contain values above \texttt{target}.
The semantic partition invariants become useful only when the boundary invariants (\textbf{I2} and \textbf{I3}) hold.

Our dependency analysis makes this semantic structure operational by testing which previously established invariants are needed for another invariant to become inductive.
In this example, the analysis identifies I1 as a prerequisite for proving the bounds invariants I2 and I3, and identifies the established bounds facts as prerequisites for proving the partition invariants I4 and I5.
The resulting graph is therefore not just a syntactic relation among clauses: it reflects that high-level partition information is supported by lower-level index-range information.
This motivates us to use the dependency graph to approximate the semantic relationships among invariants.

\begin{figure}[!t]
    \centering
    \includegraphics[width=\linewidth]{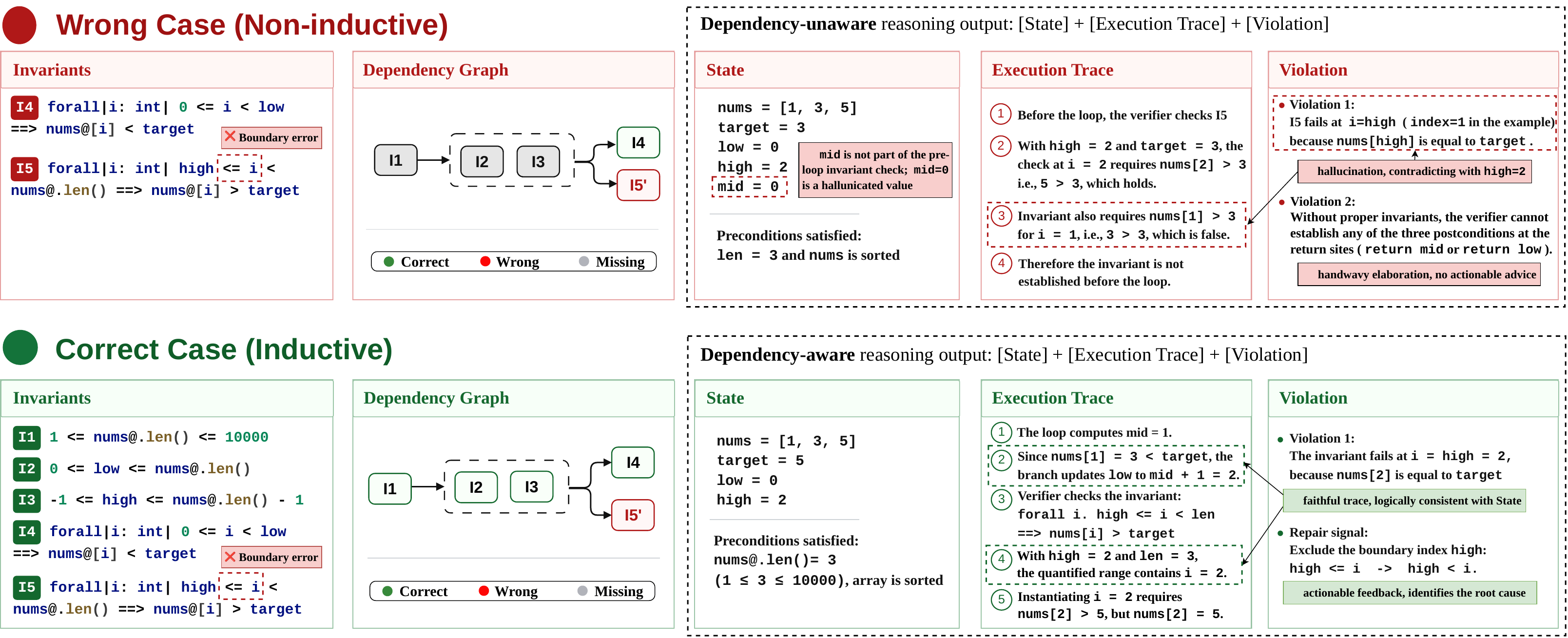}
    \vspace{-10pt}
    \caption{Comparison of dependency-unaware (top) and dependency-aware (bottom) diagnostics.}
    \label{fig:motivated_v1}
    \vspace{-10pt}
  \end{figure}

\para{Under-constrained diagnostic.}
Consider a repair attempt that includes I4 and an incorrect version of I5, but omits I1--I3.
The incorrect I5 includes the boundary index \texttt{high}:
\[
  \texttt{forall|i| high <= i < nums@.len() ==> nums@[i] > target}.
\]
Because I1--I3 are absent, verification may fail on preliminary preservation or postcondition obligations before isolating the semantic boundary error.
The annotated diagnostic in Figure~\ref{fig:motivated_v1} illustrates the failure mode: the output has a plausible \emph{State--Trace--Violation} format, but the three parts are not mutually consistent.
With \texttt{high=2} and \texttt{len=3}, the range \texttt{high <= i < len} contains only \texttt{i=2}; a trace that instantiates \texttt{i=1} is outside the quantified domain, even though \texttt{nums[1]} is not greater than \texttt{target}.
Its secondary violation is even less actionable: it broadly says that return-site postconditions cannot be proved, but does not identify a concrete failing return state or a local repair.
The diagnostic therefore points at I5 without providing a valid counterexample for the claimed quantified failure.

\para{Constrained diagnostic.}
Now suppose I1--I4 have already been established and the only remaining bug is the boundary in I5.
The same diagnostic format becomes informative because the state and trace are anchored by the prerequisite invariants.
The bottom half of Figure~\ref{fig:motivated_v1} shows the intended contrast: after one iteration with \texttt{nums=[1,3,5]}, \texttt{target=5}, \texttt{low=2}, and \texttt{high=2}, the bounds invariants justify that the quantified range is the intended in-bounds suffix and I4 rules out unrelated left-partition failures.
Now the incorrect I5 includes \texttt{i=high=2}, requiring \texttt{nums[2] > 5}, but \texttt{nums[2]=5}.
The counterexample directly pinpoints the repair solution: change \texttt{high <= i} to \texttt{high < i}.

\para{Implication for training.}
The example shows why a good counterexample is critical to proof progress.
If a model is rewarded only for final verifier success, it may spend attempts on diagnostics whose states and traces are not anchored by prerequisite invariants.
The Topological Graph Reward in Section~\ref{subsec:dep_reward} turns the dependency graph above into process feedback: it encourages repairs to make prerequisite facts available, so that counterexamples for subsequent, semantic invariants are checked in a better-constrained proof state.

\section{Methodology}
\label{sec:method}

\subsection{Problem Formulation}
\label{subsec:formal}

We formally define the problem of curriculum-based proof repair as a dependency-aware generation process.
Given a program $\mathcal{P}$ with a specification $\Phi = (P_{pre}, Q_{post})$, \ie pre-conditions and post-conditions, the task is to synthesize a proof $\Pi$ consisting of proof annotations (invariants, assertions, \etc) that verifies the program.

\para{Counterexample-guided repair.}
Let $\Pi_t$ denote the proof annotations at repair step $t$. If the proof contains a verification error $e_t$ corresponding to a failing condition, we decompose this failing constraint into an antecedent $A_t(\sigma)$ (pre-state) and a consequent $C_t(\sigma)$ (proof obligation).
A valid counterexample is a concrete program state $\sigma$ that witnesses this failure, satisfying:

\begin{equation}
\label{eq:cex}
\sigma \models A_t(\sigma) \;\wedge\; \neg\, C_t(\sigma),
\end{equation}

Rather than relying on an external SMT-based tool to statically resolve this state, our approach instructs the model to generate a structured diagnostic tuple ($\sigma$, $\tau$, $\psi$) as an intermediate reasoning step, \ie counterexample, execution trace, and violation in Figure~\ref{fig:motivated_v1}.
$\tau$ represents the specific execution trace by executing from $\sigma$, and $\psi$ is a natural-language description detailing why the obligation was falsified.

\para{Invariant dependency.}
Let $V=\{v_1,...,v_n\}$ denote the canonical invariant set extracted from a verified reference proof for program $\mathcal{P}$.

We model their prerequisite structure via a dependency graph $G=(V,E)$, where $(v_j,v_i) \in E$ represents that prerequisite $v_j$ must be established prior to inductively verifying $v_i$.
At step $t$, the objective is to synthesize $\Pi_{t+1}$ to resolve errors $e_t$.
Rather than blindly blocking state $\sigma$, the model learns the curriculum where it tends to follow the topological ordering of $G$, via reinforcement learning, as detailed in Section~\ref{subsec:dep_reward}.

\subsection{Reward Architecture}
\label{subsec:reward}

We optimize a response-level policy $\pi_\theta(y \mid x)$ with GRPO, where $x$ is the current repair context and $y$ is the complete structured response. The learning signal is a shaped reward:
\begin{equation}
\label{eq:reward}
R(x,y) =
S_{\text{safe}}(x,y)\Bigl[
G_{\text{format}}(y)\bigl(R_{\text{pass}}(x,y) + \lambda_{\text{dep}} R_{\text{dep}}(x,y)\bigr)
+ \lambda_{\text{format}} R_{\text{format}}(x,y)
\Bigr],
\end{equation}
where $S_{\text{safe}}$ is a safety gate, $R_{\text{pass}}$ is the primary task reward from verifier feedback, $R_{\text{dep}}$ is a dependency-aware shaping reward over proof structure, $R_{\text{format}}$ is an auxiliary format reward, and $G_{\text{format}}(y)$ acts as a feasibility gate that downweights malformed outputs.
The scalars $\lambda_{\text{dep}}$ and $\lambda_{\text{format}}$ are fixed reward weights; when the dependency reward is disabled, we set $\lambda_{\text{dep}} = 0$.

The verifier reward is binary: $R_{\text{pass}} = \mathbf{1}$ when Verus accepts the repaired program.
The format reward $R_{\text{format}} \in [0, 0.4]$ is an auxiliary term that rewards the presence of required structural components (XML reasoning tags and the three counterexample fields: state, execution trace, violation).
The format gate $G_{\text{format}}$ scales the primary reward according to structural completeness, \ie full score when both reasoning blocks are present, a reduced score when only one is present, and a smaller score when neither is present, preventing the mode collapse into standard Chain-of-Thought repair that bypasses the counterexample reasoning step.

The safety gate $S_{\text{safe}} \in \{0,1\}$ uses Lynette, an AST-level checker developed by~\cite{yang2024autoverus}, to check if the model only modified the proof annotations (invariants, assertions, \etc); unsafe modifications (\eg execution code) receive zero reward.

\subsection{Topological Graph Reward}
\label{subsec:dep_reward}

A core component of \tool is the \emph{Topological Graph Reward} ($R_{\text{dep}} \in [0, 1]$), which captures structural dependencies among proof obligations and encourages the model to establish prerequisite invariants before dependent ones.

\begin{definition}[Inductive Subset]
\label{def:inductive_subset}
Let $V = \{v_1, \ldots, v_n\}$ be the canonical set of loop invariants for a program.
A subset $S \subseteq V$ is \emph{inductive} if the conjunction $\bigwedge_{v \in S} v$ is (i)~established by the loop precondition, (ii)~preserved by every loop body iteration.
We write $\mathcal{I}(V)$ for the family of all inductive subsets of $V$.
\end{definition}

\begin{definition}[Invariant Dependency Graph]
\label{def:dep_graph}
For each invariant $v_i \in V$, define its \emph{dependency set} as:
\begin{equation}
\label{eq:deps}
\mathrm{deps}(v_i) \;=\; \Bigl(\,\bigcap_{\substack{S \in \mathcal{I}(V),\; v_i \in S}} S\Bigr) \setminus \{v_i\}.
\end{equation}
That is, $\mathrm{deps}(v_i)$ consists of the invariants present in \emph{every} inductive subset containing $v_i$, \ie the invariants that $v_i$ structurally depends on.
The \emph{invariant dependency graph} is $G = (V, E)$ with:
\begin{equation}
\label{eq:edge_set}
E \;=\; \bigl\{(v_j,\, v_i) \;\mid\; v_j \in \mathrm{deps}(v_i)\bigr\}.
\end{equation}
An edge $(v_j, v_i) \in E$ encodes that $v_j$ must be established before $v_i$ can be inductively verified.
\end{definition}

\begin{definition}[SCC Condensation]
\label{def:condensation}
When $G = (V, E)$ contains cycles (mutually inductive invariants), we compute its strongly connected components $\mathrm{SCC}(G) = \{S_1, \ldots, S_p\}$ via Tarjan's algorithm and form the condensation:
\begin{equation}
\hat{G} \;=\; \bigl(\hat{V},\;\hat{E}\bigr), \quad \hat{V} = \mathrm{SCC}(G), \quad
\hat{E} = \bigl\{(S_a, S_b) \;\mid\; \exists\, v_i \in S_a,\; v_j \in S_b: (v_i, v_j) \in E,\; a \neq b\bigr\}.
\end{equation}
The condensation $\hat{G}$ is a DAG that admits a topological ordering $\prec$ over $\hat{V}$, capturing dependency order represented by the universal-prerequisite graph.
\end{definition}

\para{Invariant matching.}
Given the model's generated code, we extract candidate invariants and match them against the canonical set $V$ via: (1)~exact substring matching, and (2)~fuzzy token-set matching with Jaccard similarity $J(\cdot, \cdot) \geq 0.6$.
Let $V' \subseteq V$ denote the set of matched canonical invariants.

\para{Reward computation.}
Let $E' = \{(v_j, v_i) \in E \mid v_j, v_i \in V'\}$ be the dependency edges between matched invariants, let $\hat{E}' = \{(v_j, v_i) \in E' \mid \hat{S}(v_j) \neq \hat{S}(v_i)\}$ be the inter-SCC edges, and let $\mathrm{pos}(v)$ denote the source position of invariant $v$ in the model's output.
The topological graph reward is:
\begin{equation}
\label{eq:dep_reward}
R_{\text{dep}} = \underbrace{\frac{|V'|}{|V|}}_{\alpha\;(\text{coverage})} \;\times\; \underbrace{\frac{\bigl|\{(v_j, v_i) \in \hat{E}' \mid \mathrm{pos}(v_j) < \mathrm{pos}(v_i)\}\bigr|}{|\hat{E}'|}}_{\beta\;(\text{edge satisfaction})},
\end{equation}
where $\hat{S}(v)$ denotes the SCC of $v$ in the condensation $\hat{G}$, and $\mathrm{pos}(v_j) < \mathrm{pos}(v_i)$ means the model generated $v_j$ before $v_i$.
Intra-SCC edges (mutually inductive invariants) are excluded, since no meaningful ordering exists within an SCC.
When $\hat{E}' = \emptyset$, we set $\beta = 1$.
The coverage term $\alpha$ rewards producing more canonical invariants; the edge satisfaction term $\beta$ rewards generating prerequisite invariants before dependent ones, inducing a micro-curriculum within each repair attempt.

\begin{remark}
The dependency graph $G$ and inductive subsets $\mathcal{I}(V)$ are pre-computed offline for each training instance.
In training, the reward computation reduces to invariant matching and DAG edge checking, \ie an $O(|V| + |E|)$ operation that adds negligible overhead to the GRPO training loop.

\end{remark}

\section{Evaluation}
\label{sec:evaluation}

\subsection{Experimental Setup}
\label{subsec:setup}

\textbf{Benchmarks.}
We evaluate on 471 test problems from seven Verus benchmark suites:
\textbf{VerusBench}~(26 general verification problems from AutoVerus~\citep{yang2024autoverus}),
\textbf{DafnyBench}~(17 translated Dafny problems~\citep{loughridge2024dafnybench}),
\textbf{HumanEval}~(15 algorithmic problems~\citep{chen2021evaluating}),
\textbf{LeetCode}~(5 algorithmic problems~\citep{leetcode2026}),
\textbf{VeriCogen}~(89 proof-stripped HumanEval variants from AlphaVerus~\citep{bursuc2025benchmark}),
\textbf{ObfsBench}~(266 problems with obfuscated variable names~\citep{yang2026exverus}), and
\textbf{VeruSAGE}~(53 systems-level tasks from Anvil, Mimalloc, and NrOS~\citep{yang2026verusage}).

\textbf{Models.}
All \tool variants start from QwQ-32B~\citep{qwen2025qwq32b}.
\textbf{CoT-Init} is supervised fine-tuning with chain-of-thought repair only.
\textbf{CTX-Init} uses the counterexample output format but no RL.
\textbf{CTX-Binary} applies GRPO with a binary verifier reward.
\textbf{CTX-Dep}, the full \tool model, applies GRPO with the Topological Graph Reward.
We compare against four prompted cloud LM baselines: Sonnet-4.5~\citep{anthropic2025sonnet45}, GPT-4o~\citep{openai2024gpt4o}, Qwen3-Coder~\citep{cao2026qwen3}, and DeepSeek-V4 (using the DeepSeek-V4-Flash model).

\textbf{Protocol and metrics.}
The standard repair setting uses iterative verifier feedback for up to 10 steps.
At each repair step, the model receives the failed proof and Verus feedback.
Counterexamples are generated by the model, not provided as input.
The reference proof, its invariant set $V$, and the dependency graph $G$ are not provided during evaluation.
We report Safe-Pass@1 and Safe-Pass@3; Safe-Pass@3 counts a problem as solved if any of three independent repair chains succeeds.
A solution is \emph{safe} if Lynette confirms that executable code and user specifications were not changed.
For test-time scaling, we use 32 chains $\times$ 10 iterations for Pass@$K$ curves and 1 chain $\times$ 32 iterations for iteration-scaling curves.

\textbf{Training data.}
The SFT dataset contains 461 proof repair trajectories from 249 source programs.
To construct each trajectory, we provide the teacher model, Sonnet 4.6, with the failed proof, the corresponding verifier error, and the verified reference program.
The teacher uses the reference proof to generate intermediate reasoning that explains how to repair the failed proof.
This reasoning consists of a \texttt{<think>} block, a \texttt{<counterexample>} block with \texttt{[State]}, \texttt{[Execution Trace]}, and \texttt{[Violation]} fields, and a \texttt{<reasoning>} block containing an explicit repair plan.
We extract the generated reasoning and append the verified reference program unchanged to form the target response for SFT.
RL uses the same 249 source programs, which do not overlap with the test problems.

\textbf{Training details.}
SFT uses DeepSpeed ZeRO-3 on 4 NVIDIA L40S GPUs with learning rate $2\times10^{-5}$, batch size 32, maximum sequence length 8{,}192, and 2 epochs.
RL uses GRPO with FSDP2, 4-way Ulysses sequence parallelism, vLLM rollouts at temperature 0.7 and top-$p=0.9$, 8 rollouts per prompt, learning rate $2\times10^{-6}$, no KL penalty, PPO clip ratio 0.28, and 2 epochs over 249 training programs.

\subsection{Main Results}
\label{subsec:main_results}

\begin{table*}[t]
    \centering
    \newcolumntype{Y}{>{\raggedleft\arraybackslash}X}
    
    \caption{Overall verification repair safe pass rates. Each cell reports Safe-Pass@$k$ (\%) across seven benchmarks. Cloud LLMs use iterative repair with verifier feedback; our models use 3 chains $\times$ 10 iterations. A solution is \emph{safe} if Lynette confirms only ghost (proof) code changed.}
    \label{tab:main_results}
    \small
    \setlength{\tabcolsep}{2pt} 
    \renewcommand{\arraystretch}{1.1}

    \begin{tabular}{l cc cc cc cc cc cc cc}
    \toprule
    \textbf{Approach} & \multicolumn{2}{c}{\textbf{VerusBench}} & \multicolumn{2}{c}{\textbf{DafnyBench}} & \multicolumn{2}{c}{\textbf{HumanEval}} & \multicolumn{2}{c}{\textbf{Leetcode}} & \multicolumn{2}{c}{\textbf{VeriCogen}} & \multicolumn{2}{c}{\textbf{ObfsBench}} & \multicolumn{2}{c}{\textbf{VeruSAGE}} \\
     & \scriptsize $k=1$ & \scriptsize $k=3$ & \scriptsize $k=1$ & \scriptsize $k=3$ & \scriptsize $k=1$ & \scriptsize $k=3$ & \scriptsize $k=1$ & \scriptsize $k=3$ & \scriptsize $k=1$ & \scriptsize $k=3$ & \scriptsize $k=1$ & \scriptsize $k=3$ & \scriptsize $k=1$ & \scriptsize $k=3$ \\
    \midrule
    
    \rowcolor{gray!10}
    \multicolumn{15}{l}{\textit{Cloud LLMs}} \\
    Sonnet-4.5  & 73.1 & 84.6 & 70.6 & 76.5 & 13.3 & 13.3 & 40.0 & 40.0 & 26.7 & 30.0 & 62.0 & 75.2 & 75.5 & 84.9 \\
    GPT-4o      & 42.3 & 42.3 & 76.5 & 76.5 & 6.7  & 13.3 & 0.0  & 0.0  & 40.4 & 41.4 & 21.8 & 22.9 & 17.0 & 24.5 \\
    Qwen3-Coder & 34.6 & 34.6 & 52.9 & 52.9 & 6.7  & 6.7  & 0.0  & 0.0  & 20.0 & 25.6 & 14.3 & 18.8 & 45.3 & 64.2 \\
    DeepSeek-V4 & 23.1 & 38.5 & 47.1 & 82.4 & 13.3 & 20.0 & 0.0  & 0.0  & 28.9 & 36.7 & 12.8 & 29.3 & 58.5 & 66.0 \\
    
    \midrule
    \rowcolor{gray!10}
    \multicolumn{15}{l}{\textit{External proof-repair systems}} \\
    AutoVerus   & 38.5 & 50.0 & 58.8 & 70.6 & 6.7  & 6.7  & 20.0 & 20.0 & 22.5 & 28.1 & 13.2 & 23.7 & 7.5  & 9.4  \\
    
    \midrule
    \rowcolor{gray!10}
    \multicolumn{15}{l}{\textit{Ours (QwQ-32B)}} \\
    CoT-Init    & 46.2 & 50.0 & 82.4 & 100  & 20.0 & 26.7 & 0.0  & 0.0  & 38.2 & 48.3 & 42.9 & 60.9 & 15.1 & 24.5 \\
    CTX-Init    & 88.5 & 92.3 & 88.2 & 100  & 26.7 & 26.7 & 20.0 & 40.0 & 48.3 & 61.8 & 55.3 & 73.3 & 9.4  & 22.6 \\
    CTX-Binary  & 88.5 & 96.2 & 94.1 & 100  & 26.7 & 26.7 & 20.0 & 40.0 & 44.9 & 60.7 & 58.6 & 75.9 & 24.5 & 30.2 \\
    \textbf{CTX-Dep} & \textbf{88.5} & \textbf{88.5} & \textbf{100} & \textbf{100} & \textbf{20.0} & \textbf{33.3} & \textbf{20.0} & \textbf{20.0} & \textbf{49.4} & \textbf{65.2} & \textbf{69.9} & \textbf{80.1} & \textbf{30.2} & \textbf{43.4} \\
    \bottomrule
    \end{tabular}
\end{table*}

Table~\ref{tab:main_results} shows that \tool achieves the strongest average Safe-Pass@1 / Safe-Pass@3, 61.6\% / 72.2\%.
This is 5.7 / 6.3 points above Sonnet-4.5, the strongest cloud baseline by average, and 7.9 / 4.3 points above CTX-Binary, the strongest non-dependency 32B variant.
The clearest suite-level gain over cloud baselines is on ObfsBench, where \tool reaches 69.9\% Safe-Pass@1 and 80.1\% Safe-Pass@3, improving over Sonnet-4.5 by 7.9 and 4.9 points.
On VeruSAGE, closed cloud models remain stronger, but \tool still improves over the other QwQ-32B variants, suggesting that dependency-aware RL helps the open model without eliminating all model-capacity gaps.

\subsection{Ablation Study}
\label{subsec:ablation}

Table~\ref{tab:main_results} also isolates the effect of counterexample formatting and dependency-aware RL.
Moving from CoT-Init to CTX-Init raises weighted safe Pass@1 from 39.3\% to 50.5\%, a gain of 11.2 points.
This supports the core premise that asking the model to expose a concrete failing state before repair is more useful than unstructured reasoning.
The improvement is especially large on VerusBench (46.2\% to 88.5\%) and ObfsBench (42.9\% to 55.3\%), indicating that the format helps even before RL.

Binary RL improves CTX-Init from 50.5\% to 53.7\%, but its gains are uneven.
It improves VeruSAGE from 9.4\% to 24.5\%, yet degrades VeriCogen from 48.3\% to 44.9\%.
This pattern is consistent with the sparsity problem discussed in Section~\ref{subsec:dep_reward}: verifier success is useful, but binary reward does not distinguish whether a repair trajectory is building the proof structure in a robust order.

The full dependency reward produces the largest gain, improving weighted safe Pass@1 to 61.6\% and Pass@3 to 72.2\%.
It is the best variant on DafnyBench, VeriCogen, ObfsBench, VeruSAGE, and the weighted average.
The largest absolute gain over CTX-Binary is on ObfsBench (+11.3 Pass@1 points), where obfuscated identifiers reduce the usefulness of memorized templates and make structural proof ordering more important.

\subsection{Experiment 3: Counterexample Utility}
\label{subsec:cex_utility}

\begin{wraptable}{r}{0.56\textwidth}
\vspace{-20pt}
\centering
\caption{Counterexample usability on 118 hard failed states. Cells report frozen-repair success; colored deltas are relative changes over CoT-Init.}
\label{tab:exp3-two-repairers}
\small
\setlength{\tabcolsep}{2pt}
\begin{tabular}{@{}lccc@{}}
\toprule
CEX source & DeepSeek & Sonnet & Qwen \\
\midrule
CoT-Init & 11.9\% & 31.4\% & 13.6\% \\
CTX-Init & 15.3\% \textcolor{teal}{\scriptsize(+28.6\%)} & 32.2\% \textcolor{teal}{\scriptsize(+2.7\%)} & 12.7\% \textcolor{red}{\scriptsize(-6.2\%)} \\
CTX-Binary & 19.5\% \textcolor{teal}{\scriptsize(+64.3\%)} & 33.1\% \textcolor{teal}{\scriptsize(+5.4\%)} & 15.3\% \textcolor{teal}{\scriptsize(+12.5\%)} \\
\textbf{CTX-Dep} & \textbf{19.5\%} \textcolor{teal}{\scriptsize(+64.3\%)} & \textbf{39.0\%} \textcolor{teal}{\scriptsize(+24.3\%)} & \textbf{15.3\%} \textcolor{teal}{\scriptsize(+12.5\%)} \\
\bottomrule
\end{tabular}
\vspace{-10pt}
\end{wraptable}

Table~\ref{tab:exp3-two-repairers} shows how much generated counterexamples help repair models other than the generator itself.
We freeze 118 hard failed states from unverified repair attempts.
For each state, we keep the buggy program and verifier error fixed, replace only the counterexample source, and ask three independent repairers (DeepSeek-V4, Sonnet-4.5, and Qwen3-Coder) to produce a repair.
CTX-Dep counterexamples yield the best Sonnet repair rate, 39.0\%, a 24.3\% relative improvement over CoT-Init diagnostics.
DeepSeek-V4 and Qwen3-Coder also improve over CoT-Init when given CTX-Dep diagnostics.
The result supports a conservative claim: the learned diagnostics need not be independently formal certificates, but they carry actionable information that transfers to separate repair policies.

\subsection{Test-Time Scaling}
\label{subsec:tts}

\begin{figure}[t]
\centering
\includegraphics[width=0.8\textwidth]{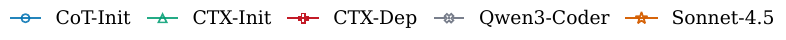}
\vspace{2pt}

\begin{minipage}[t]{0.48\textwidth}
\centering
\includegraphics[width=\linewidth]{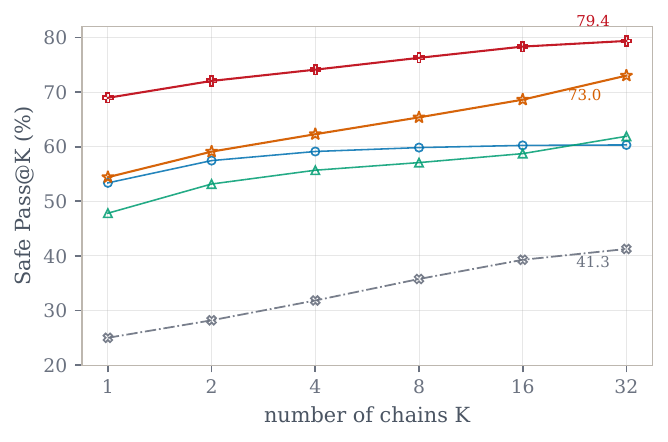}
\vspace{-5pt}
{\small(a) Sample scaling}
\end{minipage}
\begin{minipage}[t]{0.48\textwidth}
\centering
\includegraphics[width=\linewidth]{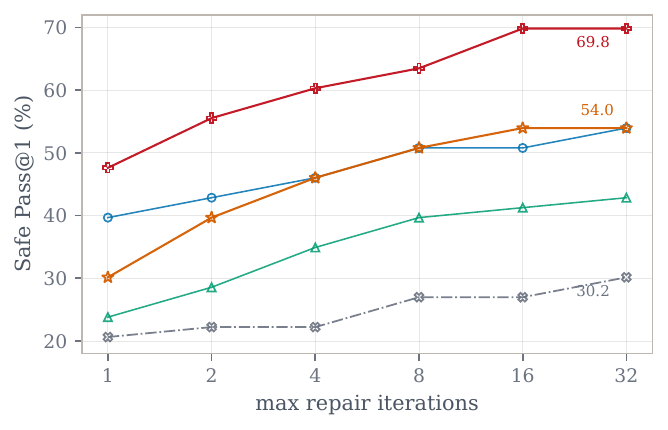}
\vspace{-5pt}
{\small(b) Iteration scaling}
\end{minipage}
\caption{Safe test-time scaling. The sample-scaling panel reports Safe Pass@$K$ as the number of chains increases, while the iteration-scaling panel reports Safe Pass@1 as the repair budget increases. }
\label{fig:safe_pass}
\vspace{-10pt}
\end{figure}

Figure~\ref{fig:safe_pass} reports test-time scaling on the 63-problem validation set drawn from VerusBench, DafnyBench, HumanEval, and LeetCode.
CTX-Dep scales best with additional test-time compute, reaching 79.4\% Safe Pass@32 in sample scaling and 69.8\% Safe Pass@1 with 32 repair iterations.
The gains over CTX-Init, CoT-Init, and Sonnet-4.5 show that dependency-aware diagnostics improve scaling with both independent samples and iterative repair budgets.

\subsection{Case Study}
\label{subsec:case_study}

We examine a maximum-profit task from \texttt{Leetcode/121-BestTimeSell} to illustrate how a generated counterexample localizes an invariant error.
The function scans a price sequence, maintaining the minimum price and maximum profit over the processed prefix, but the buggy invariant requires \texttt{max\_pro} to bound profits over the entire array.
Verus reports that the invariant is not satisfied before the loop.

Figure~\ref{fig:case-study-full} makes the scope error explicit: with \texttt{price=[5,3,7]}, \texttt{i=0}, and \texttt{max\_pro=0}, the pair \texttt{j=0}, \texttt{k=2} reduces the invariant to the false inequality \texttt{7-5 <= 0}.
The counterexample therefore pinpoints a local correction: restrict the quantified range to the processed prefix, \texttt{0 <= j < k < i}.

\begin{figure}[htbp]
\centering
\begin{casebox}{Full-array claim used as a prefix invariant}
\caseheader{Leetcode/121-BestTimeSell}
{Compute the maximum profit from one buy-sell pair in a price sequence.}

\setlength{\caseleftw}{0.485\linewidth}%
\setlength{\caserightw}{0.485\linewidth}%
\begin{lrbox}{\caseleft}\begin{minipage}[t]{\caseleftw}\setlength{\parskip}{0pt}%
\casehead{Buggy code.}
\begin{lstlisting}[style=caseverus,style=casealign,language=Rust]
pub fn max_profit(price: Vec<i32>) -> (res: i32)
    requires
        1 <= price.len() <= 100000,
        forall|i: int| 0 <= i < price.len()
            ==> 0 <= price[i] <= 10000,
    ensures
        forall|j: int, k: int| 0 <= j < k < price.len()
            ==> price[k] - price[j] <= res,
{   
    let mut max_pro = 0;
    let mut min_price = i32::MAX;
    for i in 0..price.len()
        invariant
            forall|j: int| 0 <= j < i
                ==> price[j] >= min_price,
            i > 0 ==> exists|j: int| 0 <= j < i
                && price[j] == min_price,
            // BUG: full-array range; should be k < i.
            forall|j: int, k: int|
                0 <= j < k < price.len()
                ==> price[k] - price[j] <= max_pro,
    {
        min_price = min_i32(min_price, price[i]);
        max_pro = max_i32(max_pro,
            price[i] - min_price);
    } 
    max_pro
}
\end{lstlisting}
\end{minipage}\end{lrbox}%
\begin{lrbox}{\caserighttop}\begin{minipage}[t]{\caserightw}\setlength{\parskip}{0pt}%
\casehead{Verus error.}
\begin{lstlisting}[style=caseerror,style=casealign]
error: invariant not satisfied before loop
  --> best_time_sell_v2.rs:37:13
   |
37 | forall|j: int, k: int|
   | ^^^^^^^^^^^^^^^^^^^^^^^
\end{lstlisting}
\vspace{6pt}
\casehead{Counterexample.}
\end{minipage}\end{lrbox}%
\begin{lrbox}{\caserightbot}\begin{minipage}[t]{\caserightw}\scriptsize
The last invariant talks about all future days. It should only summarize the
processed prefix: \texttt{0 <= j < k < i}.
\end{minipage}\end{lrbox}%
\noindent\usebox{\caseleft}\hfill
\begin{minipage}[t]{\caserightw}\setlength{\parskip}{0pt}%
\usebox{\caserighttop}\par\vspace{-0.92pt}\nointerlineskip%
\newcommand{\casectxtext}{%
\textbf{[State]} \texttt{price = [5,3,7]}. Before the first loop iteration,
\texttt{i = 0}, \texttt{max\_pro = 0}, and \texttt{min\_price = i32::MAX}.
All input bounds hold.

\casesep
\textbf{[Execution Trace]} Verus checks the loop invariants at entry. The
buggy invariant requires every pair in the whole array to have profit at
most \texttt{max\_pro}. Instantiate it with \texttt{j = 0}, \texttt{k = 2}.
Then \texttt{price[2] - price[0] = 7 - 5 = 2}.

\casesep
\textbf{[Violation]} The invariant demands \texttt{2 <= max\_pro = 0}, which is
false. The counterexample identifies the quantifier range as the root cause:
before processing any day, the proof can only make claims about the empty prefix
\texttt{0..i}, not about the full input sequence.}%
\newcommand{\casectxfont}{\scriptsize}%
\setlength{\casegap}{0pt}%
\begin{lrbox}{\casectx}\begin{minipage}[t]{\dimexpr\caserightw-6.8pt\relax}%
\casectxfont\setlength{\parskip}{0pt}\casectxtext\end{minipage}\end{lrbox}%
\setlength{\casectxht}{\dimexpr\casetotal{\caseleft}-\casetotal{\caserighttop}-\casetotal{\caserightbot}-3.08pt\relax}%
\setlength{\casegap}{\dimexpr(\casectxht-6.65pt-\casetotal{\casectx}-8pt)/2\relax}%
\begin{ctxbox}[before skip=0pt,after skip=0pt,valign=top,sharp corners,boxrule=0.4pt,boxsep=0pt,left=3pt,right=3pt,top=5.85pt,bottom=5.85pt,height=\casectxht,fontupper=\casectxfont]
\setlength{\parskip}{0pt}\casectxtext
\end{ctxbox}
\vspace{4pt}\nointerlineskip
\usebox{\caserightbot}
\end{minipage}
\end{casebox}

\caption{Maximum-profit case study: buggy proof, Verus feedback, and generated counterexample.}
\label{fig:case-study-full}
\end{figure}

\section{Related Work}
\label{sec:related}

\para{LM-Based formal verification.}
Recent efforts have significantly lowered the barrier to formal verification using LMs, across both mathematical theorem proving~\citep{polu2020gpt, yang2024leandojo, lample2022hypertree} and program verification~\citep{first2023baldur, sun2024clover, he2026stepwise, zhong2025rag}.
In Verus proof generation, prior works primarily focused on encoding heuristics and repair strategies via either engineering prompts~\cite{yang2024autoverus,sun2026veristruct}, or constructing large-scale synthetic training
sets~\cite{Chen2025Automated}.
However, a common limitation across these systems is their reliance on the traditional iterative-refinement paradigm, heavily depending on coarse, abstract feedback from the underlying SMT solvers.
Such feedback often fails to isolate the root cause of verification failures, frequently causing LM-based repair to degenerate into inefficient trial-and-error generation.
Our work advances this paradigm by moving beyond verifier feedback; we train the model to internalize diagnostic reasoning, integrating actionable, concrete counterexamples directly into the repair policy.

\para{Counterexample-guided feedback for proof repair.}
Traditional formal methods have long relied on counterexamples, \ie concrete execution traces that witness proof failures, as the key actionable feedback to iteratively refine invariants and generate proofs~\citep{clarke2000counterexample,bradley2012understanding}.
Recent works have begun to explore LM-driven counterexample generation to sidestep the challenge of reconstructing counterexamples from low-level SMT solver models~\citep{yang2026exverus,li2026learning}.
Nonetheless, these existing methods often demand strict formal validation of the generated counterexamples, which is often intractable, especially for complex proofs where there are multiple counterexamples to different missing/wrong invariants.
Following the philosophy of reasoning over counterexamples, \tool aims to generate high-quality counterexamples, but in a principled way, \ie internalizing the capability of generating high-quality counterexamples following the topological dependency graph.
Rather than searching for arbitrary counterexamples for a proof, \tool trains the model to produce counterexamples whose diagnostic content is useful for successful downstream repair.

\para{Reinforcement learning and curriculum learning.}
Reinforcement learning with verifiable rewards has shown strong results in theorem proving~\citep{alphaproof2024, xin2024deepseekprover, wang2025kimina} and code generation~\citep{le2022coderl, liu2023rltf}, but optimizing complex verification tasks remains challenging due to reward sparsity and long proof trajectories.
Curriculum learning~\citep{bengio2009curriculum, portelas2020automatic,wang2021survey} mitigates such sparsity by structuring learning from easy to hard; traditionally this operates across episodes as a macro-curriculum over training data.
More recently, work on long-horizon reasoning has begun exploring intra-episode micro-curricula that structure the generation process within a single rollout~\citep{dou2024stepcoder, zhang2026sortedrl, shen2025thinking}.
These approaches define curricula through external heuristics (e.g., sequence length, task decomposition), whereas our Topological Graph Reward extracts an intrinsic micro-curriculum directly from the natural dependency DAG among Verus proof obligations, rewarding the model for establishing prerequisite invariants before more complex ones within each repair trajectory.

\section{Discussion and Conclusion}
\para{Limitations.}
First, the dependency-aware micro-curriculum is fundamentally targeting loop invariants.
While invariants are often a key bottleneck of formal verification, this focus restricts the framework's immediate generalization to other critical verification structures, \eg assertions. 
Second, the validity of the generated counterexamples is not formally guaranteed.
Because exact reachability checking is fundamentally challenging and often infeasible~\citep{yang2026exverus}, the model treats the counterexample $(\sigma, \tau, \psi)$ strictly as an intermediate reasoning step optimized for downstream repair success rather than absolute mathematical validity.
Consequently, the model may occasionally hallucinate imprecise execution traces that provide ambiguous guidance.

\para{Conclusion.} We presented \tool, a 32B-parameter framework that transforms automated formal verification from a chaotic, unordered trial-and-error process into a dependency-aware generation pipeline.
By training a response-level policy via GRPO, our model learns to synthesize structured, intermediate counterexamples in a principled way and incrementally repair the proofs.
The Topological Graph Reward extracts a natural micro-curriculum directly from the DAG of proof obligations, explicitly incentivizing the model to establish foundational constraints before writing dependent invariants.
Our extensive evaluations demonstrate that synergistically coupling counterexample-guided feedback with topological curriculum learning significantly mitigates the sparsity of verifier rewards and outperforms the baselines.

\section*{Ethics Statement}
This work develops AI methods for automated formal verification, which aims to improve the reliability and correctness of software systems.
We do not foresee direct negative societal impacts.
All benchmarks used are publicly available or synthesized from open-source codebases.
Our models are trained on code verification tasks and are not designed for generating harmful content.

\begin{ack}
We thank Shan Lu for insightful discussions and the anonymous reviewers for their helpful feedback.
Results presented in this paper were obtained using the Chameleon testbed~\citep{keahey2020lessons} supported by the National Science Foundation.
\end{ack}

\bibliography{main}
\bibliographystyle{format/colm2026_conference}

\appendix
\section{Inductive Subset Enumeration Algorithm}
\label{app:algorithm}

\subsection{Problem Formulation and Computational Complexity}

Let $\mathcal{V} = \{v_1, \ldots, v_n\}$ be the set of loop invariant clauses in a verified Verus program.
A subset $S \subseteq \mathcal{V}$ is \emph{inductive} if the conjunction of its clauses holds at loop entry and is preserved by each loop iteration.
If $S$ is inductive, no strict superset of $S$ can be a minimal inductive subset.
The dependency graph $G_\mathcal{V}$ is derived from the \emph{minimal inductive subsets} $\mathcal{I}_{\min}$ (Definitions~\ref{def:inductive_subset}, \ref{def:dep_graph}, and~\ref{def:condensation} in the main text).

\paragraph{Hardness of complete enumeration.}
Enumerating \emph{all} minimal inductive subsets is closely related to---and, if inductiveness were monotone under taking supersets, equivalent to---the \emph{Hypergraph Transversal Enumeration} problem (Trans-Enum)~\citep{eiter1995transversal}.
Formally, given a hypergraph $H = ([n], \mathcal{E})$, a \emph{minimal transversal} is a minimal set $T \subseteq [n]$ that intersects every hyperedge; enumerating all minimal transversals is precisely our problem when each hyperedge encodes a ``necessary support'' constraint for inductiveness.

This problem occupies a rare position in the complexity landscape~\citep{eiter1995transversal}.
Despite 30+ years of effort, it is not known to be solvable in polynomial time for general hypergraphs.
At the same time, it is unlikely to be coNP-hard: because it admits a quasi-polynomial-time algorithm, coNP-hardness would place all of coNP in quasi-polynomial time.
The best known algorithm runs in quasi-polynomial time $N^{O(\log N / \log \log N)}$~\citep{fredman1996dualization}, where $N$ is the combined size of the input and output.
In this sense it resembles Graph Isomorphism: no polynomial-time algorithm is known, yet it is not believed to be NP-hard.
The connection to formal verification is not coincidental: the same structure arises as \emph{Minimal Inductive Validity Core} (MIVC) enumeration~\citep{ghassabani2016mivc}, where one seeks all minimal subsets of a transition system sufficient to prove a safety property.

The intractability of exact enumeration motivates a practical approximation: collect a sufficient sample of inductive subsets within a fixed verifier call budget $B$, yielding an approximate $\mathcal{I}_{\min}$ that is exact for the vast majority of programs and near-exact for the rest.

\subsection{Level-wise BFS Enumeration Algorithm}

Given that exact enumeration is intractable in general, we design an algorithm that (i) is \emph{correct} when the budget suffices to exhaust the search space, and (ii) degrades gracefully to a well-structured approximation when the budget is exceeded.
The algorithm exploits the fact that the minimal inductive subsets form an antichain, via a level-wise breadth-first search:
subsets are enumerated in increasing cardinality order $k=1,2,\ldots,n$, and any superset of an already-found inductive set is pruned without a verifier call.

\begin{remark}[Correctness under mutual inductiveness]
Some invariant groups are \emph{mutually inductive}: no individual clause $\{v_i\}$ verifies on its own, but the group $S = \{v_i, v_j, \ldots\}$ does (these form SCCs in $G_{\mathcal{V}}$).
A seed-based approach that initializes from singletons would produce an empty queue for such groups and miss them entirely.
The level-wise algorithm below handles this correctly because it tests subsets of every size $k = 1, 2, \ldots, n$ (subject to the budget), so a mutually inductive group is found at level $|S|$ even when none of its singletons is inductive.
\end{remark}

\begin{algorithm}[H]
\caption{Level-wise BFS Enumeration of Minimal Inductive Subsets}
\label{alg:levelwise}
\begin{algorithmic}[1]
\Require Invariant set $\mathcal{V} = \{v_1,\ldots,v_n\}$, verifier oracle $\textsc{Verify}(S) \to \{\texttt{true},\texttt{false}\}$, budget $B$
\Ensure Family $\mathcal{I}_{\min}$ of minimal inductive subsets
\State $\mathcal{I}_{\min} \leftarrow \emptyset$; \quad $\text{Visited} \leftarrow \emptyset$; \quad $\text{calls} \leftarrow 0$
\For{$k = 1, 2, \ldots, n$}   \Comment{level = subset size}
  \For{each $S \in \binom{\mathcal{V}}{k}$ in lexicographic order}
    \If{$S \in \text{Visited}$} \textbf{continue} \EndIf
    \State $\text{Visited} \leftarrow \text{Visited} \cup \{S\}$
    \If{$\exists\, T \in \mathcal{I}_{\min}: T \subseteq S$} \textbf{continue}   \Comment{antichain pruning}
    \EndIf
    \If{$\text{calls} \geq B$} \textbf{goto} \textsc{Done} \EndIf
    \State $\text{calls} \leftarrow \text{calls} + 1$
    \If{$\textsc{Verify}(S)$}
      \State $\mathcal{I}_{\min} \leftarrow \mathcal{I}_{\min} \cup \{S\}$
    \EndIf
  \EndFor
  \If{$\text{calls} \geq B$} \textbf{break} \EndIf
\EndFor
\State \textsc{Done:} \Return $\mathcal{I}_{\min}$
\end{algorithmic}
\end{algorithm}

\subsection{Complexity Analysis}

\paragraph{Verifier call budget.}
Each subset $S$ is tested at most once (tracked in $\text{Visited}$).
The antichain pruning step (line~7 of Algorithm~1) skips any superset of an already-found inductive set: once a minimal inductive subset $T$ is found, none of its $2^{n-|T|}-1$ strict supersets can be minimal, so they are pruned without further verifier calls. In the best case (every singleton is self-inductive), the algorithm makes only $n$ calls at level $k=1$ and prunes all remaining $2^n-n-1$ subsets. In the worst case (no pruning applies), it tests all $2^n-1$ nonempty subsets, the same as naive enumeration; in practice, pruning removes most supersets at higher levels (Table~3).
For programs with $n > 8$ (2\% of our dataset), we apply a budget cap of $B = 256$ calls.

\paragraph{Empirical preprocessing statistics.}
We ran the level-wise BFS enumeration offline on all 531 verified programs across five benchmark datasets (Table~\ref{tab:preprocess}).
The median program has $n=3$ invariants; 98\% have $n \leq 8$. During actual training, we only use dependency graphs from 249 training samples.

\begin{table}[h]
\centering
\small
\caption{Preprocessing statistics per dataset. ``Calls'' = actual verifier invocations (total). ``Naive'' = calls without the budget cap ($2^n - 1$ per file, summed). ``Reduction'' = Naive / Calls. ``Est.\ time'' assumes 0.42\,s per Verus call (measured mean on our hardware; range 0.25--0.71\,s).}
\label{tab:preprocess}
\begin{tabular}{lrrrrrr}
\toprule
Dataset & Files & $n$ (mean) & Calls (total) & Naive (total) & Reduction & Est.\ time \\
\midrule
Dafnybench            &  55 & 1.9 &    205 &     205 &  1.0$\times$ &  1.4\,min \\
HumanEval-Alphaverus  &  47 & 3.4 &  1,135 &   2,669 &  2.4$\times$ &  7.9\,min \\
LeetCode              &  23 & 6.6 &  2,817 &  72,955 & 25.9$\times$ & 19.7\,min \\
Obfuscated-VerusBench & 266 & 3.0 &  3,498 &   3,498 &  1.0$\times$ & 24.5\,min \\
cleaned-VerusBench    & 140 & 2.7 &  1,561 &   1,816 &  1.2$\times$ & 10.9\,min \\
\midrule
\textbf{Total}        & \textbf{531} & \textbf{3.0} & \textbf{9,216} & \textbf{81,143} & \textbf{8.8$\times$} & \textbf{$\approx$64\,min} \\
\bottomrule
\end{tabular}
\end{table}

This concretizes the approximation motivated in \S\ref{app:algorithm}.1:
for the 98\% of programs with $n \leq 8$, the full search space ($2^n - 1 \leq 255$ subsets) is exhausted within budget $B=256$, so $\mathcal{I}_{\min}$ is \emph{exact}---equivalent to the intractable complete enumeration, achieved cheaply because our benchmark programs are small.
For the remaining 2\% ($n > 8$, all from LeetCode, $n \leq 16$), the budget cap triggers the approximation: the returned $\mathcal{I}_{\min}$ contains the minimal inductive subsets found within $B$ calls, which are guaranteed to be \emph{locally} minimal (no proper subset was found inductive) but may not be globally minimal.
The level-wise traversal order ensures that small subsets are tested first, so the approximation is biased toward finding small inductive sets---the most informative ones for dependency graph construction.
These 9 files constitute a negligible fraction of the training data, and the overall reduction in verifier calls is $8.8\times$ (9,216 actual vs.\ 81,143 naive), with LeetCode benefiting most ($25.9\times$).

\paragraph{Dependency graph construction and amortization.}
Given $\mathcal{I}_{\min}$, the dependency edges $E$ are computed in $O(|\mathcal{I}_{\min}| \cdot n^2)$ time (Definition~\ref{def:dep_graph}), and SCC condensation via Tarjan's algorithm runs in $O(n + |E|)$---both polynomial in $n$.
The entire offline preprocessing---9,216 verifier calls across 531 programs---completes in approximately 64 minutes on a single machine (measured mean Verus call time: 0.42\,s; range 0.25--0.71\,s across benchmarks), and is performed once before training begins.
At training time, the reward function performs only a dictionary lookup ($O(1)$) plus topological order evaluation ($O(n + |E|)$), adding negligible overhead to the RL rollout.

\section{Qualitative Counterexample Case Studies}
\label{app:case-studies}

We include three representative examples where the generated counterexample
context directly identifies the proof obligation that should be repaired. In
each example, the display keeps the source-level artifact fixed: the left panel
contains the task and buggy proof fragment, while the right panel contains the
raw verifier message and the generated \texttt{[State]},
\texttt{[Execution Trace]}, and \texttt{[Violation]} fields. The accompanying
text below analyzes why the raw error message is insufficient and how the
counterexample makes the repair target explicit.

\paragraph{Assertion placement.}
The first case illustrates a failure mode where the relevant counterexample is
an intermediate program state rather than a loop-entry or loop-exit state. The
task is to move elements from \texttt{shrink} into \texttt{grow} while
preserving the sum of their lengths. The loop invariant is semantically correct
at the loop boundary, but the assertion is placed after \texttt{shrink.pop()}
and before \texttt{grow.push(i)}, where the length sum is temporarily one
smaller. The verifier message reports only an assertion failure at the source
line, so it does not distinguish a wrong invariant from an assertion placed at
the wrong program point. The generated counterexample makes this distinction
explicit by tracing the execution to the precise state where the assertion is
evaluated.

\begin{casebox}{Case Study A: Assertion Checked at the Wrong Program Point}
\caseheader{Dafnybench/program\_422\_count\_less\_than}
{Move elements from \texttt{shrink} to \texttt{grow} while preserving the total length.}

\noindent
\begin{minipage}[t]{0.485\textwidth}
\textbf{\small Buggy code.}
\begin{lstlisting}[style=caseverus,language=Rust]
fn count_less_than(numbers: Vec<u32>, threshold: u32)
    -> (count: u32)
    ensures count == numbers.len() as u32,
{
    let mut shrink = numbers.clone();
    let mut grow: Vec<u32> = Vec::new();

    while shrink.len() > 0
        invariant
            grow.len() + shrink.len() == numbers.len(),
        decreases shrink.len(),
    {
        let i = shrink.pop().unwrap();
        // BUG: checked before grow.push(i).
        assert(grow.len() + shrink.len()
            == numbers.len());
        grow.push(i);
    }
    assert(numbers.len() == grow.len());
    grow.len() as u32
}
\end{lstlisting}
\scriptsize
The assertion is checked after \texttt{pop} and before the compensating
\texttt{push}; the loop invariant only holds at loop boundaries.
\end{minipage}\hfill
\begin{minipage}[t]{0.485\textwidth}
\textbf{\small Verus error.}
\begin{lstlisting}[style=caseerror]
error: assertion failed
  --> program_422_count_less_than.rs:19:16
   |
19 | assert(grow.len() + shrink.len() == numbers.len());
   |        ^^^^^^^^^^^^^^^^^^^^^^^^^^^^^^^^^^^^^^^^^^
\end{lstlisting}

\textbf{\small Counterexample.}
\begin{ctxbox}
\textbf{[State]} \texttt{numbers = [1,2,3]}, \texttt{threshold = 5}. At loop entry:
\texttt{grow = []}, \texttt{shrink = [1,2,3]}; hence
\texttt{grow.len() + shrink.len() = 0 + 3 = 3}.

\smallskip
\textbf{[Execution Trace]} Execute the first loop iteration. The statement
\texttt{shrink.pop()} removes one element, so \texttt{shrink = [1,2]} while
\texttt{grow} is still empty. The assertion is checked immediately at this
intermediate point, before \texttt{grow.push(i)}.

\smallskip
\textbf{[Violation]} At the assertion site,
\texttt{grow.len() + shrink.len() = 0 + 2 = 2}, but
\texttt{numbers.len() = 3}. The invariant is re-established only after the
subsequent push; it cannot justify an assertion placed between the two updates.
\end{ctxbox}
\end{minipage}
\end{casebox}

\paragraph{Prefix-scoped semantic invariant.}
The second case demonstrates how counterexample analysis identifies the scope of
a semantic loop invariant. The function computes the maximum profit over a
price sequence, but during the loop \texttt{max\_pro} only summarizes the prefix
processed so far. The buggy proof copies the final full-array property into the
loop invariant, requiring \texttt{max\_pro} to bound profits involving future
prices even before the first iteration. The verifier reports that the invariant
is not satisfied before the loop, but the message does not reveal which
quantifier range is too strong. The generated counterexample instantiates the
full-array claim with \texttt{prices = [5,3,7]}, \texttt{j = 0}, and
\texttt{k = 2}, yielding the false obligation \texttt{7 - 5 <= 0}. This turns a
coarse invariant failure into a localized repair target: replace the full-array
range with the processed-prefix range \texttt{k < i}.

\begin{casebox}{Case Study B: Full-array claim used as a prefix invariant}
\caseheader{Leetcode/121-BestTimeSell}
{Compute the maximum profit from one buy-sell pair in a price sequence.}

\setlength{\caseleftw}{0.485\linewidth}%
\setlength{\caserightw}{0.485\linewidth}%
\begin{lrbox}{\caseleft}\begin{minipage}[t]{\caseleftw}\setlength{\parskip}{0pt}%
\casehead{Buggy code.}
\begin{lstlisting}[style=caseverus,style=casealign,language=Rust]
pub fn max_profit(price: Vec<i32>) -> (res: i32)
    requires
        1 <= price.len() <= 100000,
        forall|i: int| 0 <= i < price.len()
            ==> 0 <= price[i] <= 10000,
    ensures
        forall|j: int, k: int| 0 <= j < k < price.len()
            ==> price[k] - price[j] <= res,
{   
    let mut max_pro = 0;
    let mut min_price = i32::MAX;
    for i in 0..price.len()
        invariant
            forall|j: int| 0 <= j < i
                ==> price[j] >= min_price,
            i > 0 ==> exists|j: int| 0 <= j < i
                && price[j] == min_price,
            // BUG: full-array range; should be k < i.
            forall|j: int, k: int|
                0 <= j < k < price.len()
                ==> price[k] - price[j] <= max_pro,
    {
        min_price = min_i32(min_price, price[i]);
        max_pro = max_i32(max_pro,
            price[i] - min_price);
    } 
    max_pro
}
\end{lstlisting}
\end{minipage}\end{lrbox}%
\begin{lrbox}{\caserighttop}\begin{minipage}[t]{\caserightw}\setlength{\parskip}{0pt}%
\casehead{Verus error.}
\begin{lstlisting}[style=caseerror,style=casealign]
error: invariant not satisfied before loop
  --> best_time_sell_v2.rs:37:13
   |
37 | forall|j: int, k: int|
   | ^^^^^^^^^^^^^^^^^^^^^^^
\end{lstlisting}
\vspace{6pt}
\casehead{Counterexample.}
\end{minipage}\end{lrbox}%
\begin{lrbox}{\caserightbot}\begin{minipage}[t]{\caserightw}\scriptsize
The last invariant talks about all future days. It should only summarize the
processed prefix: \texttt{0 <= j < k < i}.
\end{minipage}\end{lrbox}%
\noindent\usebox{\caseleft}\hfill
\begin{minipage}[t]{\caserightw}\setlength{\parskip}{0pt}%
\usebox{\caserighttop}\par\vspace{-0.92pt}\nointerlineskip%
\newcommand{\casectxtext}{%
\textbf{[State]} \texttt{price = [5,3,7]}. Before the first loop iteration,
\texttt{i = 0}, \texttt{max\_pro = 0}, and \texttt{min\_price = i32::MAX}.
All input bounds hold.

\casesep
\textbf{[Execution Trace]} Verus checks the loop invariants at entry. The
buggy invariant requires every pair in the whole array to have profit at
most \texttt{max\_pro}. Instantiate it with \texttt{j = 0}, \texttt{k = 2}.
Then \texttt{price[2] - price[0] = 7 - 5 = 2}.

\casesep
\textbf{[Violation]} The invariant demands \texttt{2 <= max\_pro = 0}, which is
false. The counterexample identifies the quantifier range as the root cause:
before processing any day, the proof can only make claims about the empty prefix
\texttt{0..i}, not about the full input sequence.}%
\newcommand{\casectxfont}{\scriptsize}%
\setlength{\casegap}{0pt}%
\begin{lrbox}{\casectx}\begin{minipage}[t]{\dimexpr\caserightw-6.8pt\relax}%
\casectxfont\setlength{\parskip}{0pt}\casectxtext\end{minipage}\end{lrbox}%
\setlength{\casectxht}{\dimexpr\casetotal{\caseleft}-\casetotal{\caserighttop}-\casetotal{\caserightbot}-3.08pt\relax}%
\setlength{\casegap}{\dimexpr(\casectxht-6.65pt-\casetotal{\casectx}-8pt)/2\relax}%
\begin{ctxbox}[before skip=0pt,after skip=0pt,valign=top,sharp corners,boxrule=0.4pt,boxsep=0pt,left=3pt,right=3pt,top=5.85pt,bottom=5.85pt,height=\casectxht,fontupper=\casectxfont]
\setlength{\parskip}{0pt}\casectxtext
\end{ctxbox}
\vspace{4pt}\nointerlineskip
\usebox{\caserightbot}
\end{minipage}
\end{casebox}

\paragraph{Arithmetic-safety context.}
The third case shows that a counterexample can diagnose proof-context loss
rather than a concrete runtime overflow. The function preconditions already
state that each square and cube of an input element is within the \texttt{i32}
range, but the loop invariant only records the output length and the values
already pushed. Consequently, the verifier reports a possible overflow at the
multiplication site, without indicating that the missing ingredient is the
arithmetic-bound precondition inside the loop context. The generated
counterexample keeps the execution state concrete and benign, then explains why
the verifier still rejects it: at the current index \texttt{i}, the loop
context does not contain the bound needed to prove
\texttt{nums[i] * nums[i] * nums[i]} safe.

\begin{casebox}{Case Study C: Missing Arithmetic-Safety Fact in the Loop Context}
\caseheader{cleaned-verusbench/task\_id\_447}
{Return a vector containing the cube of each input element.}

\noindent
\begin{minipage}[t]{0.485\textwidth}
\textbf{\small Buggy code.}
\begin{lstlisting}[style=caseverus,language=Rust]
fn cube_element(nums: &Vec<i32>) -> (cubed: Vec<i32>)
    requires
        forall|k: int| 0 <= k < nums.len()
            ==> i32::MIN <= nums[k] * nums[k]
                <= i32::MAX,
        forall|k: int| 0 <= k < nums.len()
            ==> i32::MIN <= nums[k] * nums[k] * nums[k]
                <= i32::MAX,
    ensures
        forall|i: int| 0 <= i < nums.len()
            ==> cubed[i] == nums[i] * nums[i] * nums[i],
{
    let mut cubed_array: Vec<i32> = Vec::new();
    let mut i = 0;
    while i < nums.len()
        invariant
            i <= nums.len(),
            cubed_array.len() == i,
            forall|k: int| 0 <= k < i
                ==> cubed_array[k] == nums[k] * nums[k] * nums[k],
        decreases nums.len() - i,
    {
        // BUG: loop invariants omit cube bounds.
        cubed_array.push(nums[i] * nums[i] * nums[i]);
        i += 1;
    }
    cubed_array
}
\end{lstlisting}
\scriptsize
The function precondition has the needed arithmetic bounds, but the loop
invariant does not carry them into the loop body.
\end{minipage}\hfill
\begin{minipage}[t]{0.485\textwidth}
\textbf{\small Verus error.}
\begin{lstlisting}[style=caseerror]
error: possible arithmetic underflow/overflow
  --> task_id_447.rs:40:26
   |
40 | cubed_array.push(nums[i] * nums[i] * nums[i]);
   |                  ^^^^^^^^^^^^^^^^^^^^^^^^^^^
\end{lstlisting}

\textbf{\small Counterexample.}
\begin{ctxbox}
\textbf{[State]} \texttt{nums = [10,20,30]}. At the first iteration,
\texttt{i = 0}, \texttt{cubed\_array.len() = 0}, and all function preconditions
are satisfied.

\smallskip
\textbf{[Execution Trace]} The loop guard gives \texttt{i < nums.len()}. The
program then evaluates \texttt{nums[i] * nums[i] * nums[i]}. Although the
function precondition bounds every square and cube, the loop invariant only
states the index bound, result length, and previously computed elements.

\smallskip
\textbf{[Violation]} Verus cannot prove that the current multiplication is
within the \texttt{i32} range from the loop context alone. The repair is to
carry the arithmetic-bound facts as loop invariants, so the verifier can
instantiate them at the current index \texttt{i}.
\end{ctxbox}
\end{minipage}
\end{casebox}

\section{Additional Results}

\subsection{Micro- and Macro-Averaged Safe-Pass}
\label{app:micro_macro}

The reported ``weighted average'' is a micro-average over all 471 evaluation problems, so each problem contributes equally.
It does not use manually selected dataset weights.
The macro-average first computes the performance on each benchmark suite and then averages the seven suite-level results, giving each suite equal weight regardless of its size.
Both averages use the Safe-Pass@1 and Safe-Pass@3 definitions in Section~\ref{subsec:setup}.

As shown in Table~\ref{tab:micro_macro_appendix}, CTX-Dep remains the best of the three methods under macro averaging.
It outperforms Sonnet-4.5 by 2.4/3.7 percentage points and CTX-Binary by 3.0/0.1 points in Safe-Pass@1/@3.

\begin{table}[t]
\centering
\small
\caption{Micro- and macro-averaged Safe-Pass (\%) over the 471 evaluation problems in seven benchmark suites. Micro averaging gives each problem equal weight; macro averaging gives each suite equal weight.}
\label{tab:micro_macro_appendix}
\begin{tabular}{lcccc}
\toprule
& \multicolumn{2}{c}{Micro average} & \multicolumn{2}{c}{Macro average} \\
Method & Safe-Pass@1 & Safe-Pass@3 & Safe-Pass@1 & Safe-Pass@3 \\
\midrule
Sonnet-4.5 & 55.9 & 65.9 & 51.6 & 57.8 \\
CTX-Binary & 53.7 & 67.9 & 51.0 & 61.4 \\
CTX-Dep & 61.6 & 72.2 & 54.0 & 61.5 \\
\bottomrule
\end{tabular}
\end{table}

\subsection{CTX-Binary under iteration scaling.}
We compare CTX-Binary with CTX-Dep under the same iteration-scaling setup, including the 63-task subset, prompts, and inference budgets (Section~\ref{subsec:tts}).
Each task uses one repair chain with up to 32 iterations.
In Table~\ref{tab:tts_binary_appendix}, CTX-Dep remains ahead at every iteration budget beyond the first, reaching 69.8\% Safe-Pass@1 at 32 iterations compared with 68.3\% for CTX-Binary.
The largest gap appears at two iterations (55.6\% vs.\ 49.2\%), suggesting that CTX-Dep improves more rapidly during the early repair iterations.

\begin{table}[t]
\centering
\small
\caption{Safe-Pass@1 (\%) under iteration scaling on the 63-task subset. Differences are CTX-Dep minus CTX-Binary in percentage points, computed from the displayed percentages.}
\label{tab:tts_binary_appendix}
\begin{tabular}{rccc}
\toprule
Iterations & CTX-Binary & CTX-Dep & Difference \\
\midrule
1  & 47.6 & 47.6 & 0.0 \\
2  & 49.2 & 55.6 & +6.4 \\
4  & 57.1 & 60.3 & +3.2 \\
8  & 61.9 & 63.5 & +1.6 \\
16 & 68.3 & 69.8 & +1.5 \\
32 & 68.3 & 69.8 & +1.5 \\
\bottomrule
\end{tabular}
\end{table}

\subsection{Proof Diversity}
\label{app:proof_diversity}

We compare the diversity of the proofs generated by CTX-Binary and CTX-Dep at evaluation time.
This analysis examines whether training with the dependency reward causes CTX-Dep to generate only a narrow set of proof forms.
To focus on proof annotations, we define each repaired program's proof-only signature as the whitespace-normalized concatenation of all ghost-code spans extracted by Lynette, excluding executable and other non-proof text.
Each task is evaluated with three independent repair chains.
The paired comparison includes 249 tasks from the 471-task evaluation for which both models produce at least two safe verified repairs.

We report three metrics: Ghost unique is the mean number of distinct proof-only signatures per task; Ghost rate divides this count by the number of available safe verified repairs; and Ghost distance is the mean pairwise token-Jaccard distance between signatures.
Higher values indicate greater proof diversity.
Ghost rate and Ghost distance are computed per task and then averaged over the paired eligible tasks; Ghost distance includes pairs with identical signatures.

\begin{table}[t]
\centering
\small
\caption{Diversity of proof-only signatures among paired safe verified repairs from the 471-task evaluation. Each cell reports CTX-Binary / CTX-Dep; each metric is averaged over the paired eligible tasks in its benchmark group.}
\label{tab:proof_diversity_appendix}
\begin{tabular}{lccc}
\toprule
Benchmark & Ghost unique & Ghost rate & Ghost distance \\
\midrule
Four in-distribution benchmarks & 2.524 / 2.357 & 84.9\% / 79.4\% & 0.058 / 0.040 \\
VeriCogen & 2.522 / 2.217 & 94.2\% / 88.4\% & 0.388 / 0.316 \\
ObfsBench & 2.618 / 2.640 & 96.9\% / 93.9\% & 0.072 / 0.067 \\
VeruSAGE & 1.667 / 1.667 & 72.2\% / 66.7\% & 0.220 / 0.202 \\
\bottomrule
\end{tabular}
\end{table}

The four in-distribution benchmarks comprise VerusBench, DafnyBench, HumanEval, and LeetCode.
CTX-Dep continues to generate multiple distinct proof-only signatures across all benchmark groups.
Its Ghost unique, Ghost rate, and Ghost distance are lower than those of CTX-Binary on most groups, while its Ghost-unique score is slightly higher on ObfsBench and equal on VeruSAGE.
Taken together with the overall results, these findings indicate a favorable diversity--performance tradeoff: CTX-Dep is slightly less diverse by these measures, but it continues to generate multiple proof forms across all benchmark groups while improving micro-averaged Safe-Pass@1/@3 over CTX-Binary by 7.9/4.3 percentage points (Table~\ref{tab:micro_macro_appendix}).

\subsection{Invariant Coverage and Dependency Ordering}
\label{app:invariant_ordering}

We further examine whether the dependency reward changes the order
of matched invariants in the generated proofs.
The analysis uses final outputs from the 63-task iteration-scaling evaluation.
This post-hoc analysis retains 24 paired tasks for which both model outputs pass Verus, the task has a nonempty reference dependency graph, and each output matches both endpoint invariants of at least one inter-SCC dependency edge. 
The diagnostic matcher preserves exact substring matches and then applies one-to-one token-set Jaccard matching to the remaining invariants at a threshold of 0.6.
These are post-hoc diagnostic scores, not dependency rewards recorded during training.

The metrics use the coverage and edge-satisfaction definitions in
Section~\ref{subsec:dep_reward} (Equation~\eqref{eq:dep_reward}),
evaluated with this diagnostic matcher.
Mean invariant coverage is the mean value of $\alpha = |V'|/|V|$, where $V'$ is the set of canonical invariants matched in the generated proof.
Mean edge satisfaction is the mean value of
\[
\beta = \frac{\bigl|\{(v_j,v_i)\in\hat{E}' \mid \mathrm{pos}(v_j)<\mathrm{pos}(v_i)\}\bigr|}{|\hat{E}'|},
\]
namely the fraction of considered inter-SCC dependency edges for which the prerequisite invariant appears before the invariant that depends on it.
Here $\hat{E}'$ includes only edges whose endpoints are both matched, and intra-SCC edges are excluded.
Mean invariant coverage and mean edge satisfaction average the per-output values over the same 24 tasks, with one output per model per task.
Full edge satisfaction counts tasks with $\beta=1$, with 24 as the denominator; the mean topology score averages $\alpha\beta$ over these tasks.

\begin{table}[t]
\centering
\small
\caption{Invariant coverage and dependency ordering on 24 paired tasks accepted by Verus, using the post-hoc diagnostic matcher. Differences are CTX-Dep minus CTX-Binary, computed before rounding; ``pp'' denotes percentage points.}
\label{tab:invariant_ordering_appendix}
\begin{tabular}{lccc}
\toprule
Metric & CTX-Binary & CTX-Dep & Difference \\
\midrule
Mean invariant coverage, $\alpha$ & 98.6\% & 98.5\% & $-0.1$ pp \\
Mean edge satisfaction, $\beta$ & 43.8\% & 95.8\% & $+52.1$ pp \\
Tasks with full edge satisfaction, $\beta=1$ & 10/24 (41.7\%) & 23/24 (95.8\%) & $+54.2$ pp \\
Mean topology score, $\alpha\beta$ & 0.424 & 0.943 & $+0.519$ \\
\bottomrule
\end{tabular}
\end{table}

CTX-Binary and CTX-Dep have nearly identical mean invariant coverage, 98.6\% and 98.5\%, respectively.
The main difference is therefore not the number of canonical invariants matched in the generated proofs.
Instead, CTX-Dep increases mean edge satisfaction from 43.8\% to 95.8\%, and increases the fraction of tasks with $\beta=1$ from 41.7\% to 95.8\%.
This means that, among the matched invariants and considered dependency edges, CTX-Dep is substantially more likely to generate the prerequisite invariant before the invariant that depends on it.

\section{Prompt Templates}
\label{app:prompts}

We document the prompt templates used during RL training and iterative test-time
evaluation.
Both templates elicit the same structured output format: a chain-of-thought
reasoning block $\langle$\texttt{think}$\rangle$, a semi-formal counterexample
block $\langle$\texttt{counterexample}$\rangle$ with three mandatory subsections
(\texttt{[State]}, \texttt{[Execution Trace]}, \texttt{[Violation]}), a repair
rationale block $\langle$\texttt{reasoning}$\rangle$, and a single
\texttt{rust} code block containing the repaired program.

\subsection{RL Training Prompt}
\label{app:prompts:rl}

During GRPO training, each rollout is conditioned on a fixed system message
that establishes the model's role and a per-problem user message that
supplies the buggy program and the Verus verifier output.

\begin{promptbox}[System Message]
\begin{verbatim}
You are an expert in Verus, a formal verification framework for Rust.
You fix buggy Verus programs by analyzing verification errors,
constructing concrete counterexamples, and writing repaired code
that passes the verifier.

Common Verus fix patterns:
- Missing decreases clause: add `decreases <expr>` to while/for loops
- Postcondition not satisfied: add loop invariants that mirror the
  postcondition
- Precondition not satisfied: add invariants proving bounds
  (e.g., index < len)
- Arithmetic overflow: add bounds invariants or use checked arithmetic
- Recommendation not met: add `assert` to guide the prover
\end{verbatim}
\end{promptbox}

\begin{promptbox}[User Message]
\begin{verbatim}
Fix the following buggy Verus Rust program so it passes the verifier.

Respond with these sections in order:
1. <counterexample> with [State], [Execution Trace], [Violation]
2. <reasoning> with a concise repair plan
3. A single ```rust ... ``` block with the complete fixed program

---
Benchmark: {benchmark}
Problem:   {problem}

[Buggy Code]
```rust
{buggy_code}
```

[Verifier Error]
{verifier_error}
\end{verbatim}
\end{promptbox}

\subsection{Iterative Repair Prompt (Refinement Turn)}
\label{app:prompts:refine}

During test-time scaling, the model is queried for up to $T$ iterations.
For iteration $t \geq 2$, a refinement turn is appended after the original
prompt, supplying the previous repair attempt and the verifier's updated
error log.
This enables the model to construct a fresh counterexample and revise its
repair strategy based on the remaining verification failures.

\begin{promptbox}[Refinement Turn (appended after the original prompt, iteration $t \geq 2$)]
\begin{verbatim}
{original_prompt}

---
# Previous Attempt
The code generated in the previous turn was:
```rust
{previous_code}
```

# Verifier Output
The Verus verifier reported the following errors:
{verifier_output}

# Task
Analyze the remaining verification errors. Respond with:
1. <counterexample> with [State], [Execution Trace], [Violation]
2. <reasoning> with a concise repair plan
3. A single ```rust ... ``` block with the complete fixed program.
\end{verbatim}
\end{promptbox}

\newpage
\section*{NeurIPS Paper Checklist}

The checklist is designed to encourage best practices for responsible machine learning research, addressing issues of reproducibility, transparency, research ethics, and societal impact. Do not remove the checklist: {\bf The papers not including the checklist will be desk rejected.} The checklist should follow the references and follow the (optional) supplemental material.  The checklist does NOT count towards the page
limit. 

Please read the checklist guidelines carefully for information on how to answer these questions. For each question in the checklist:
\begin{itemize}
    \item You should answer \answerYes{}, \answerNo{}, or \answerNA{}.
    \item \answerNA{} means either that the question is Not Applicable for that particular paper or the relevant information is Not Available.
    \item Please provide a short (1--2 sentence) justification right after your answer (even for \answerNA). 
\end{itemize}

{\bf The checklist answers are an integral part of your paper submission.} They are visible to the reviewers, area chairs, senior area chairs, and ethics reviewers. You will also be asked to include it (after eventual revisions) with the final version of your paper, and its final version will be published with the paper.

The reviewers of your paper will be asked to use the checklist as one of the factors in their evaluation. While \answerYes{} is generally preferable to \answerNo{}, it is perfectly acceptable to answer \answerNo{} provided a proper justification is given (e.g., error bars are not reported because it would be too computationally expensive'' or ``we were unable to find the license for the dataset we used''). In general, answering \answerNo{} or \answerNA{} is not grounds for rejection. While the questions are phrased in a binary way, we acknowledge that the true answer is often more nuanced, so please just use your best judgment and write a justification to elaborate. All supporting evidence can appear either in the main paper or the supplemental material, provided in appendix. If you answer \answerYes{} to a question, in the justification please point to the section(s) where related material for the question can be found.

\begin{enumerate}

\item {\bf Claims}
    \item[] Question: Do the main claims made in the abstract and introduction accurately reflect the paper's contributions and scope?
    \item[] Answer: \answerYes{} %
    \item[] Justification: The abstract and introduction accurately reflect the paper's contributions and scope.

\item {\bf Limitations}
    \item[] Question: Does the paper discuss the limitations of the work performed by the authors?
    \item[] Answer: \answerYes{} %
    \item[] Justification: The limitation discussion can be found in the Discussion and Conclusion section.

\item {\bf Theory assumptions and proofs}
    \item[] Question: For each theoretical result, does the paper provide the full set of assumptions and a complete (and correct) proof?
    \item[] Answer: \answerNA{} %
    \item[] Justification: The theoretical formulations are defined in Section~\ref{sec:method}.

    \item {\bf Experimental result reproducibility}
    \item[] Question: Does the paper fully disclose all the information needed to reproduce the main experimental results of the paper to the extent that it affects the main claims and/or conclusions of the paper (regardless of whether the code and data are provided or not)?
    \item[] Answer: \answerYes{} %
    \item[] Justification: Section~\ref{subsec:setup} describes the training and evaluation protocols and links to the public code and model checkpoint. The repository provides configurations and instructions for running the evaluation.

\item {\bf Open access to data and code}
    \item[] Question: Does the paper provide open access to the data and code, with sufficient instructions to faithfully reproduce the main experimental results, as described in supplemental material?
    \item[] Answer: \answerYes{} %
    \item[] Justification: The training and evaluation code, benchmark and instructions are available in the GitHub repository linked in Introduction.

\item {\bf Experimental setting/details}
    \item[] Question: Does the paper specify all the training and test details (e.g., data splits, hyperparameters, how they were chosen, type of optimizer) necessary to understand the results?
    \item[] Answer: \answerYes{} %
    \item[] Justification: Section~\ref{subsec:setup} describes the benchmarks and data splits, training hyperparameters, and evaluation protocol. Appendix~\ref{app:prompts} provides the prompt templates.

\item {\bf Experiment statistical significance}
    \item[] Question: Does the paper report error bars suitably and correctly defined or other appropriate information about the statistical significance of the experiments?
    \item[] Answer: \answerNo{} %
    \item[] Justification: The experiment details are in Section~\ref{sec:evaluation}. We do not report error bars or multiple training seeds because each RL run is computationally expensive; Safe-Pass@3 over three independent repair chains partially reflects sampling variability.

\item {\bf Experiments compute resources}
    \item[] Question: For each experiment, does the paper provide sufficient information on the computer resources (type of compute workers, memory, time of execution) needed to reproduce the experiments?
    \item[] Answer: \answerYes{} %
    \item[] Justification: The experiment details are in Section~\ref{sec:evaluation}.
    
\item {\bf Code of ethics}
    \item[] Question: Does the research conducted in the paper conform, in every respect, with the NeurIPS Code of Ethics \url{https://neurips.cc/public/EthicsGuidelines}?
    \item[] Answer: \answerYes{} %
    \item[] Justification: The research conforms with the Code of Ethics.

\item {\bf Broader impacts}
    \item[] Question: Does the paper discuss both potential positive societal impacts and negative societal impacts of the work performed?
    \item[] Answer: \answerYes{} %
    \item[] Justification: The broader impacts are discussed in the supplemental materials.
    
\item {\bf Safeguards}
    \item[] Question: Does the paper describe safeguards that have been put in place for responsible release of data or models that have a high risk for misuse (e.g., pre-trained language models, image generators, or scraped datasets)?
    \item[] Answer: \answerYes{} %
    \item[] Justification: The experiment details are in Section~\ref{sec:evaluation}. Verus verifier execution is safe and has no risk.

\item {\bf Licenses for existing assets}
    \item[] Question: Are the creators or original owners of assets (e.g., code, data, models), used in the paper, properly credited and are the license and terms of use explicitly mentioned and properly respected?
    \item[] Answer: \answerYes{} %
    \item[] Justification: The experiment details are in Section~\ref{sec:evaluation}. All used datasets are public.

\item {\bf New assets}
    \item[] Question: Are new assets introduced in the paper well documented and is the documentation provided alongside the assets?
    \item[] Answer: \answerYes{} %
    \item[] Justification: The trained model and data are in supplemental materials.

\item {\bf Crowdsourcing and research with human subjects}
    \item[] Question: For crowdsourcing experiments and research with human subjects, does the paper include the full text of instructions given to participants and screenshots, if applicable, as well as details about compensation (if any)? 
    \item[] Answer: \answerNA{} %
    \item[] Justification: We did not do crowdsource experiments.

\item {\bf Institutional review board (IRB) approvals or equivalent for research with human subjects}
    \item[] Question: Does the paper describe potential risks incurred by study participants, whether such risks were disclosed to the subjects, and whether Institutional Review Board (IRB) approvals (or an equivalent approval/review based on the requirements of your country or institution) were obtained?
    \item[] Answer: \answerNA{} %
    \item[] Justification: We did not do crowdsource experiments.

\item {\bf Declaration of LM usage}
    \item[] Question: Does the paper describe the usage of LMs if it is an important, original, or non-standard component of the core methods in this research? Note that if the LM is used only for writing, editing, or formatting purposes and does \emph{not} impact the core methodology, scientific rigor, or originality of the research, declaration is not required.
    \item[] Answer: \answerYes{} %
    \item[] Justification: We use a teacher LM to construct training trajectories and train TopoVerus to generate counterexamples and propose proof repairs. Sections~\ref{sec:method} and~\ref{subsec:setup} describe the method and training setup, and Appendix~\ref{app:prompts} provides the prompt templates.
    \item[] Guidelines:
    \begin{itemize}
        \item The answer \answerNA{} means that the core method development in this research does not involve LMs as any important, original, or non-standard components.
        \item Please refer to our LM policy in the NeurIPS handbook for what should or should not be described.
    \end{itemize}

\end{enumerate}

\end{document}